# Charge-density-wave quantum critical point under pressure in 2$H$-TaSe$_2$

Yuliia Tymoshenko,[1] Amir-Abbas Haghighirad,[1] Rolf Heid,[1] Tom Lacmann,[1] Alsu Ivashko,[1] Adrian Merritt,[2] Xingchen Shen,[1,3] Michael Merz,[1] Gaston Garbarino,[4] Luigi Paolasini,[4] Alexei Bosak,[4] Florian K. Diekmann,[5,6] Kai Rossnagel,[5,6,7] Stephan Rosenkranz,[8] Ayman H. Said,[9] and Frank Weber[1,†]

[1] Institute for Quantum Materials and Technologies, Karlsruhe Institute of Technology, 76131 Karlsruhe, Germany
[2] Physikalisches Institut, Karlsruhe Institute of Technology, 76131 Karlsruhe, Germany
[3] Laboratoire de Cristallographie et Sciences des Matériaux (CRISMAT), CNRS, ENSICAEN, Caen 14000, France
[4] ESRF, The European Synchrotron, 71 Avenue des Martyrs, CS40220, 38043 Grenoble Cedex 9, France
[5] Institute of Experimental and Applied Physics and KiNSIS, Kiel University, 24098 Kiel, Germany
[6] Ruprecht Haensel Laboratory, Kiel University, 24098 Kiel, Germany
[7] Ruprecht Haensel Laboratory, Deutsches Elektronen-Synchrotron DESY, 22607 Hamburg, Germany
[8] Material Science Division, Argonne National Laboratory, Lemont (IL), USA
[9] Advanced Photon Source, Argonne National Laboratory, Lemont (IL), USA

Suppressing of an ordered state that competes with superconductivity is one route to enhance superconducting transition temperatures. Whereas the effect of suppressing magnetic states is still not fully understood, materials featuring charge-density waves and superconductivity offer a clearer scenario as both states can be associated with electron-phonon coupling. Metallic transition-metal dichalcogenides are prime examples for such intertwined electron-phonon-driven phases, yet, various compounds do not show the expected interrelation or feature additional mechanisms which makes an unambiguous interpretation difficult. Here, we report high-pressure X-ray diffraction and inelastic X-ray scattering measurements of the prototypical transition-metal dichalcogenide 2$H$-TaSe$_2$ and determine the evolution of the charge-density-wave state and its lattice dynamics up to and beyond its suppression at the critical pressure $p_c$ = 19.9(1) GPa and at low temperatures. The high quality of our data allows the full refinement of the commensurate charge-density-wave superstructure at low pressure and we find the quantum critical point of the charge-density-wave to be in close vicinity to the reported maximum superconducting transition temperature $T_{sc}$ = 8.2 K. *Ab-initio* calculations corroborate that 2$H$-TaSe$_2$ is a reference example of order-suppressed enhanced superconductivity and can serve as a textbook case to investigate superconductivity near a charge-density-wave quantum critical point.

## Introduction

Superconductivity is often found in the vicinity of suppressed ordered states. Indeed, a correlation between the suppression of an ordered state, e.g., by chemical substitution or pressure, and a maximum of the superconducting transition temperature $T_{sc}$ is known in various materials[1-4]. Quantum-critical fluctuations of the suppressed ordered state are assumed to play a central role for emergent superconductivity and a dome-shaped superconducting state, e.g., upon doping or pressure, is often considered a hallmark of unconventional superconductivity[2,5-7]. Among others, charge-density-wave (CDW) order, a periodic modulation of the charge carrier density accompanied by a periodic lattice distortion, often coexists or competes with superconductivity, *e.g.*, in copper-oxide superconductors[2,8-10] and transition-metal-based Kagome metals[11-17]. On the other hand, classic CDW materials such as the layered transition-metal dichalcogenides (TMD) continue to provide a rich playground for emergent physics including chiral CDW[18-21], excitonic insulators[22,23], dimensionality-dependent correlated electronic phases[22-25] as well as intriguing magnetic properties[26] with potential applications in spintronic devices[27].

In metallic TMDs, CDW order stabilized by electron-phonon coupling (EPC) is widespread [28-30] and it is now evident that many TMDs feature an emergent superconducting state once CDW is sufficiently suppressed.[31-42] The generic phase diagram of such emergent superconductivity[2,5-7] (Fig. 1) shows the maximum of $T_{sc}$ at the critical point where the nearby order is suppressed to zero temperature, i.e., at its quantum critical point (QCP). It is generally thought that order-parameter fluctuations support the emergent superconducting state. For a suppressed CDW state, the CDW soft-phonon mode corresponds to zero-energy quantum critical fluctuations and might explain that $T_{sc,max}$ is located at the CDW QCP. Yet, prototypical CDW-hosting TMDs like 2$H$-NbSe$_2$ do not reflect this scenario[43-47] or feature more complex underlying microscopic mechanisms such as 1$T$-TiSe$_2$.[38,41]

Here, we employ high-pressure X-ray diffraction (XRD) and inelastic X-ray scattering (IXS) to study the evolution of CDW order and its associated soft phonon mode under pressure up to 30 GPa and down to 4 K in the TMD 2$H$-TaSe$_2$. 2$H$-TaSe$_2$ is a classic CDW compound featuring a large periodic lattice distortion[48] and a momentum-dependent energy gap in the electronic band structure in its low-temperature state[49,50]. It is a layered material [Fig. 2(a)] for which CDW order with a transition temperature $T_{ICDW}$ = 122.3 K was reported in the 1970s.[51] On cooling through $T_{ICDW}$, 2$H$-TaSe$_2$ first enters a CDW state with an incommensurate ordering wave vector $\mathbf{q}_{ICDW}$ = (0.323,0,0) [all wave vectors are given in reciprocal lattice units (r.l.u.); see Methods], which evolves on cooling and reaches the commensurate value $\mathbf{q}_{CCDW}$ = (1/3,0,0) at $T_{CCDW} \approx$ 90 K.[52] The evolution of CDW order with pressure was previously investigated by lab-based XRD up to 5 GPa [53] and resistivity [33] but the CDW QCP could not be determined. On the other hand, the superconducting transition temperature of 2$H$-TaSe$_2$ increases smoothly from $T_{sc}$ = 0.1 K at ambient pressure to $T_{sc,max}$ = 8.2 K at 23-25 GPa before it starts to decrease again.[33] Our combined XRD-IXS study supported by *ab-initio* lattice dynamical calculations demonstrates that 2$H$-TaSe$_2$ features a CDW QCP at $p_c$ = 19.9(1) GPa in vicinity of $T_{sc,max}$ and, thus, can serve as a reference example for order-suppressed enhanced superconductivity near a CDW quantum-critical point.

## Results

Our focus is on the structural evolution and, in particular, that of CDW order in 2$H$-TaSe$_2$ under pressure, addressing the question whether a CDW QCP exists close to $T_{sc,max}$. We performed synchrotron x-ray diffraction at the European Synchrotron Radiation Facility (ESRF, see Methods for more details) at pressures up to 30 GPa and temperatures down to 10 K. The setup of the high-pressure diamond-anvil cell (DAC) is sketched in Figure 2(b) where helium gas was used as pressure medium. A typical 2D slice of the data set taken at T = 40 K and p = 0.3 GPa [Fig. 2(c)] reveals the main Bragg reflections of the hexagonal lattice (large dark spots) accompanied by commensurate CDW peaks along the equivalent high-symmetry directions [100], [010] and [-110]. The CDW peaks divide the line between two hexagonal reciprocal lattice points in three equally large sections [inset in Fig. 2(c)]. Thus, in agreement with previous work [48,51-53], we find a commensurate CDW ordering wave vector, $q_{CCDW}$ = (1/3,0,0), at low temperatures and low pressures.

In total we measured more than 120 different temperature-pressure points (see Fig. S1) and found that the observed Bragg reflections can always be best indexed using the high-temperature hexagonal unit cell with space group $P6_3/mmc$ [Fig. 2(a)]. The corresponding refined structural parameters (not considering the CDW superlattice peaks) are detailed in Table 1 for selected temperatures at low and high pressures. Results for all data points are available[54] along with raw data and crystallographic information files (CIF) of the measurements highlighted in Table 1.

Furthermore, we present a full refinement of the commensurate CDW structure at T = 40 K and p = 0.3 GPa. The commensurate CDW unit cell represents a 3×3×1 supercell of the high temperature structure maintaining its hexagonal space group (see Table 1). Figure 3 shows the commensurate CDW hexagonal supercell with atoms at the positions of the undistorted high-temperature structure. Deviations of the atomic positions in the commensurate CDW state are indicated by arrows for each atomic site. The Ta displacements are strictly in the *a-b* plane whereas some Se atoms show deviations from their position in the high-temperature hexagonal unit cell also along the *c* axis. The refined distortion validates previous qualitative predictions about the structure in the commensurate CDW phase based on neutron diffraction[48] resolving a long-standing debate in 2*H*-TaSe$_2$.[55-58]

Having established the general structural properties of 2*H*-TaSe$_2$, we turn to a detailed analysis of the CDW superlattice peak. We exemplify this analysis on data taken along $Q$ = (1,0,0) – (2,0,0) and T = 40 K for pressures up to 20 GPa [Fig. 4(a)]. The pressure evolution of the superlattice peak at $Q$ ≈ (1.33,0,0) illustrates an intriguing behavior [Fig. 4(b)]: starting with the commensurate value at low pressures (blue), the CDW wave vector $q_{CDW}$ = (1/3-δ,0,0) acquires a small incommensuration δ ≈ 0.0015 r.l.u. above 1 GPa (orange) and becomes commensurate again for pressures from 5 GPa (green) to 8.6 GPa. At pressures p > 10 GPa we find a strongly incommensurate CDW order with $δ_{max}$ = 0.023 (red). Whereas the intensity of the CDW superlattice peak is fairly constant a lower pressures, it reduces strongly in the high-pressure incommensurate state indicating a strong decrease of the atomic lattice distortions associated with CDW order. At 18.2 GPa and T = 40 K (purple), we only find a broad hump of scattered intensity near $Q$ = (1.31,0,0) with a weak intensity reduced by more than two orders of magnitude. The full pressure dependence of the CDW incommensuration δ and the peak width along with refined structural parameters of the high-temperature unit cell are shown in Figures 4(c)-(f). The evolution of the CDW peak linewidth reveals a critical pressure at T = 40 K of around 18 GPa (vertical dashed line). We do not find a response in the structural parameters [Figs. 4(c,d) and Suppl. Figs. S2,S3] to the suppression of CDW order nor to the onset of the strong incommensuration. XRD measurements at T = 10 K were done for pressures up to 15 GPa and near 23 GPa where we found that the CDW order is already suppressed. We complemented our results near 20 GPa by elastic momentum scans at T = 10 K performed on the ID28 IXS spectrometer located at ESRF, France. Here, scans with zero energy transfer across the CDW superlattice peak at $Q$ = (2.64,0,1) (see Suppl. Fig. S4) define the critical pressure of the CDW order in 2*H*-TaSe$_2$ at this temperature to be between 19.7 GPa and 20 GPa, i.e. $p_c$ = 19.9(1) GPa.

Summarizing the above described results, we show the temperature-pressure CDW phase diagram of 2*H*-TaSe$_2$ in Figure 5(a) complemented by pressure-dependent superconducting transition temperatures taken from Ref. [33]. The incommensurate CDW state is suppressed continuously from 122 K at ambient pressure to zero just below 20 GPa near the maximum of $T_{sc}$.[33] We assign the discrepancy of up to 4-5 GPa between $p_c$ and the reported pressure range of $T_{sc,max}$ to the use of a different pressure medium. Freitas *et al.*[33] report $T_{CDW}$ ≈ 70 K at p = 20 GPa and the two highest values of $T_{sc}$ for 23 GPa and 27 GPa yielding an approximate average pressure of 25(2) GPa for $T_{sc,max}$. In our experiments using helium as the most hydrostatic pressure medium, $T_{CDW}$ ≈ 70 K is observed at 15.5 GPa, i.e., 4.4 GPa below the corresponding critical pressure, $p_c$ = 19.9(1) GPa [Fig. 5(a)]. Thus, our results for pressures of 15 – 20 GPa correspond to results published by Freitas et al.[33] at pressures which are 4-5 GPa higher. Therefore, we conclude that the CDW QCP of 2*H*-TaSe$_2$ is close to the pressure of T$_{sc,max}$.

The low-temperature commensurate CDW order is gradually suppressed at low-pressures though CDW order at T = 10 K stays commensurate at all investigated pressures up to 5 GPa (solid gray line, in agreement with previous reports [59]). At T = 40 K, CDW superlattice peaks are found at commensurate positions again for pressures 5 GPa ≤ p ≤ 8.6 GPa. The presence of the reentrant commensurate CDW state is in qualitative agreement with previous reports, though quantitatively we find the onset at higher pressure than reported in 1980 [53]. Again, the use of a different pressure medium, i.e., methanol-ethanol mix, and pressure calibration limited to room temperature[53] might explain such discrepancies. In regions where we did not perform a dense set of XRD measurements, we indicate the evolution of T$_{CCDW}$ at the mid-points between measurements featuring commensurate and incommensurate CDW order (gray dashed lines).

One requirement for a quantum critical point is that the suppressed phase transition is continuous, i.e., of 2$^{nd}$ order. The presence of a soft phonon mode indicates a 2$^{nd}$ order phase transition and, therefore, we employed inelastic x-ray scattering to check the evolution of the CDW soft phonon at high pressures and low temperatures.[a] The low-temperature pressure-dependent phonon energy of the same mode studied at ambient pressure[47] confirms its soft-mode character in the vicinity of p$_c$ [Fig. 5(b) and Suppl. Fig. S5] and is evidence that the suppressed phase transition remains continuous. Thus, we conclude that 2*H*-TaSe$_2$ features a CDW QCP at p$_c$ = 19.9(1) GPa.

Detailed lattice dynamical calculations can assess not only the phonon dispersions in a material but also the strength of EPC and provide an estimate of the corresponding superconducting transition temperature. In a recent publication[47], we have shown that calculations using the hexagonal high-temperature structure confirm the suppression of the CDW phonon instability at a pressure of 23 GPa. The calculated EPC, which is mostly confined to the original CDW soft phonon mode in 2*H*-TaSe$_2$, can explain superconducting transition temperatures of about 10 K. Here, we present a more detailed set of calculations, where pressure is included by using the lattice parameters determined in our XRD measurements. Figure 5(c) shows the calculated square of the CDW soft mode energy as function of pressure (grey symbols). Negative values at lower pressures indicate an imaginary frequency of the CDW soft phonon mode and, thus, the formation of a CDW superstructure. On the other hand, the positive values of the calculated square phonon energies of the CDW soft mode show that the ambient-pressure high-temperature structure [see illustration in Fig. 2(a)] is stable at low-temperature for pressures larger than p$_{c,DFPT}$ = 18.8 GPa [grey symbols in Fig. 5(c)]. Thus, the calculated critical pressure[b] of the CDW state is quite close to the observed value of p$_c$ = 19.9 GPa.

We computed the full phonon dispersion spectrum and the corresponding EPC for stable structures above p$_{c,DFPT}$ (see Fig. 6 in Ref. [47] for p = 23 GPa). To estimate T$_{sc}$ at high pressures, we solved the linearized gap equation of the Eliashberg theory on the imaginary axis[60]. This equation takes as input the Eliashberg function $\alpha^2 F(\omega)$ and the parameter µ*, which represents an effective Coulomb repulsion. Using a typical value of $\mu^* = 0.1$, we got a maximum $T_{sc,\max(DFPT)} = 11.9$ K at p$_{c,DFPT}$ = 18.8 GPa [green symbols in Fig. 5(c)]. The low-energy modes including the soft branch contribute 73% to the total coupling constant $\lambda_{\mathrm{EPC}} = 1.03$. When ignoring this contribution, $T_{sc(DFPT)}$ is zero. Including the full strength of EPC for pressures p ≥ 18.8 GPa, the pressure dependence of $T_{sc(DFPT)}$ is negative though the reduction is small, i.e., ΔT$_{sc}$ = 1.7 K for Δp = 5 GPa [green symbols in Fig. 5(c)], in reasonable agreement with the published results.[33] Therefore, our calculations support the scenario that order-parameter fluctuations represented by the CDW soft phonon mode mediate emergent superconductivity in 2*H*-TaSe$_2$.

---

[a] A full report of the (ongoing) inelastic scattering experiments will be published elsewhere.
[b] We remind that pressure in the calculations refers to the pressure at which the corresponding lattice parameters, used for the calculations, were observed.

## Discussion

Emergent superconductivity in the vicinity of an ordered state is a central topic of research in many quantum materials such as the cuprates, iron-based superconductors and Kagome metals. Particularly, the role of order-parameter fluctuations of the suppressed state is intensely investigated. Do these fluctuations promote the superconducting pairing? For the TMD 2$H$-TaSe$_2$ we can conclusively say "yes". The phase diagram shows that, in contrast to previous suggestions [33], the incommensurate CDW state is continuously suppressed and reaches zero temperature in close vicinity to $T_{sc,max}$ [see Fig. 5(a)]. The observed soft-mode behavior confirms that the transition stays 2$^{nd}$ order [see Fig. 5(b)] and, therefore, a CDW QCP in 2$H$-TaSe$_2$ exists at high pressure. The estimated $T_{sc,DFPT}$ crucially depends on the low-energy EPC. Moreover, our calculations explain the broad pressure dependence of $T_{sc}$ and show that $T_{sc,max}$ is tied to the CDW QCP [see Fig. 5(c)].

The straightforwardness of EPC by which we can link $T_{sc,max}$ to the CDW QCP in 2$H$-TaSe$_2$ contrasts with other CDW-hosting TMDs behaving differently. For instance, incommensurate CDW order in isostructural 2$H$-NbSe$_2$ is suppressed near 4.4 GPa [41] but $T_{sc,max}$ occurs close to 10 GPa [38]. Moreover, harmonic calculations overestimate the CDW critical pressure by about a factor of three [45]. In fact, the presence of a CDW QCP under pressure has been questioned altogether for 2$H$-NbSe$_2$ [41]. Similarly, 1$T$-TiSe$_2$ has been intensively studied under pressure [40,42,61-63]. The most recent work [42] points towards the presence of a CDW QCP. Yet, it also reports on significant and sudden changes in the electronic band structure at 2 GPa, just before superconductivity sets in. Thus, it is not clear if CDW order and superconductivity compete in the same way as in 2$H$-TaSe$_2$.

Researchers have investigated thin samples of TMDs down to the single-(sandwich-)layer limit. Interestingly, $T_{sc}$ is reduced in thin NbSe$_2$ [31-36] whereas it increases in TaSe$_2$ by a factor of up to ten.[64,65] Recent theoretical work[58] proposes that the CDW structure in single-layer TaSe$_2$ may be different from that originally proposed one[48]. Here, our definitive structural refinement of the bulk commensurate CDW state in agreement with the original proposal provides an important starting point to understand these contrasting properties[26,66,67] and benchmark calculations in the bulk limit.

Thus, 2$H$-TaSe$_2$ under pressure provides one of the clearest examples of order-parameter-fluctuation enhanced superconductivity at a QCP. It visualizes textbook behavior and can serve as starting point to improve our understanding of more complex CDW materials such as Kagome metals[17,68], nickel pnictides[69,70] or cuprates.[8,9,71]

For instance, the Kagome system CsV$_3$Sb$_5$ shows a roughly similar temperature-pressure phase diagram featuring CDW and superconducting states with similar maximum transition temperatures of 90 K (p = 0) and 8 K (p ≈ 2 GPa), respectively, as 2$H$-TaSe$_2$.[72] A crossover between two differently ordered CDW states on increasing pressure is discussed as origin of a peculiar double-peak structure in $T_{sc}$(p) with explicit reference to 2$H$-TaSe$_2$.[72,73] According to our results, the low-temperature state of 2$H$-TaSe$_2$ is commensurate up to a pressure of about 8.6 GPa [see Fig. 5(a)]. However, $T_{sc}$ was not measured near the commensurate to incommensurate CDW crossover which is accompanied by a large drop in the superlattice peak intensity [see Fig. 4(b)]. It will be interesting to re-examine $T_{sc}$(p) more closely since Ref. [33] reports $T_{sc}$ only for pressures below (8 GPa) and well-above (11 GPa) of 8.6 GPa. We note that experiments in CsV$_3$Sb$_5$ so far did not observe a CDW soft phonon mode[16,74,75]. Theory emphasizes anharmonic phonon properties[76], yet predicts that EPC is large enough to mediate the observed superconducting transition temperatures[77].

## Methods

***X-ray diffraction*** experiments at high-pressure and low-temperature were performed at the European Synchrotron Radiation Facility (ESRF, beamline ID15B) in a membrane-type diamond anvil cell (DAC) using the ruby fluorescence method for the pressure calibration [78]. For the experiments at the ESRF an ESRF-made Le Toullec–type CuBe-alloy DAC with diamonds with cullet diameters of 500 µm was used. A stainless-steel gasket was indented down to a thickness of about 100 µm. A 2*H*-TaSe$_2$ single crystal and one ruby were placed inside the gasket hole, and helium was used as the pressure-transmitting medium. The 2*H*-TaSe$_2$ sample was a small piece cut from the same single crystal as the samples for IXS and grown at Kiel University (see [47]). A sketch of the DAC setup are shown in Figure 2(b). For x-ray diffraction a monochromatic beam with an energy of 30.17 keV (≈0.411 Å) was focused down to 4×4 µm$^2$, and the diffracted beam was detected with an EIGER2 X 9M CdTe flat panel detector. For each dataset images within an angular range of ±32° were recorded. Each image was integrated for 0.2 s and over a 0.5° rotation. The detector position and distance (180 mm) were calibrated with silicon powder and an enstatite single-crystal standard using the DIOPTAS [79] and CrysAlis-PRO softwares [80]. CrysAlis-PRO [80] was used for cell refinement, data reduction, and the analysis of the diffraction precession images for all datasets. SHELLXS97 [81] and SHELLXL97 2014/7 [82] as well as JANA2006 [83] were used to solve the crystal structure and refinements. Crystal data and structural refinement details for selected measurements are summarized in Table 1. An extended table is available electronically[54]. The x-ray crystallographic coordinates of the crystal structures we found have been deposited at the Cambridge Crystallographic Data Centre (CCDC) under deposition numbers 2415237-2415239. Atomic coordinates and site labels were standardized using the VESTA[84] crystal structure visualization software.

**Inelastic x-ray scattering** experiments at high-pressure and low-temperature were carried out at the beamline ID28 [85] at the European Synchrotron Radiation Facility (ESRF, France). The sample was a small piece cut from the same single crystal as the samples for XRD and grown at Kiel University (see [47]). The high-pressure low-temperature setup was the same as used on high-pressure XRD on beamline ID15b at ESRF (see above). Phonon excitations measured in constant-momentum scans were approximated by damped harmonic oscillator (DHO) functions[86] convoluted with a pseudo-Voigt resolution function [full-width at half-maximum (FWHM) = 1.4 meV, Lorentz ratio = 0.74]. The resolution function was further used to approximate resolution-limited elastic scattering at zero energy transfer. Measurements were done at scattering wave vectors **Q** = **τ** - **q**, where **τ** is a reciprocal lattice vector and **q** is the reduced wave vector in the 1$^{st}$ Brillouin zone. Wave vectors are expressed in reciprocal lattice units (r.l.u.) $(2\pi/a, 2\pi/b, 2\pi/c)$ with the lattice parameters $a = b \approx 3.44$ Å and $c \approx 12.7$ Å of the high-temperature hexagonal unit cell [#194, Fig. 2(a)]. All IXS measurements were done in the Brillouin zone adjacent to **τ** = (3,0,1).

***Ab-initio*** **lattice dynamical calculations** based on density-functional-perturbation-theory (DFPT) were performed in the framework of the mixed basis pseudopotential method [42]. The exchange-correlation functional was treated in the local-density approximation (LDA). Spin-orbit interaction was taken into account consistently. More details are given in Supplemental Information of Ref. [47]. Pressure dependent properties were calculated by using corresponding lattice parameters deduced from XRD at T = 40 K. The internal parameter *z* was relaxed.

**Acknowledgements**

T. L. was supported through the Deutsche Forschungsgemeinschaft (DFG, German Research Foundation) under project 422213477 (TRR 288 project B03). A. M. was supported by BMBF under contract 05K22VK1 in the framework of ErUM-PRO. X.S. was supported by the Helmholtz-OCPC Postdoc Program. S.R. was supported by the Materials Sciences and Engineering Division, Office of Basic Energy Sciences, U.S. Department of Energy. The authors gratefully acknowledge European Synchrotron Radiation Facility (ESRF) for time on beamlines ID15b (proposal HC-5083)[87] and ID28 (proposal HC-5554).

**AUTHOR CONTRIBUTIONS**

X-ray diffraction: Y.T., A.-A.H., T.L., G.G., M. M., X.C., F.W.; Inelastic x-ray scattering: Y. ., A.M., A.I., L.P., A.B., S.R., A.H.S., F.W.; Data analysis: Y.T., A.-A.H., F.W.; Theory: R.H.; Sample growth: K.R., F.D.; Manuscript: Y.T., A.-A.H., F.W.; Project coordination: F.W.

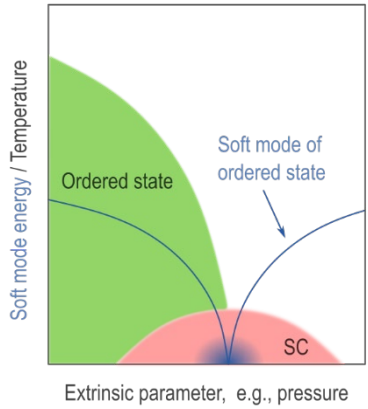

**Fig. 1.** Generic phase diagram of emergent superconductivity driven by order-parameter fluctuations in the vicinity of the suppression of an ordered state. The blue line indicates the pressure dependence of the ordered-state's soft mode at low temperatures.

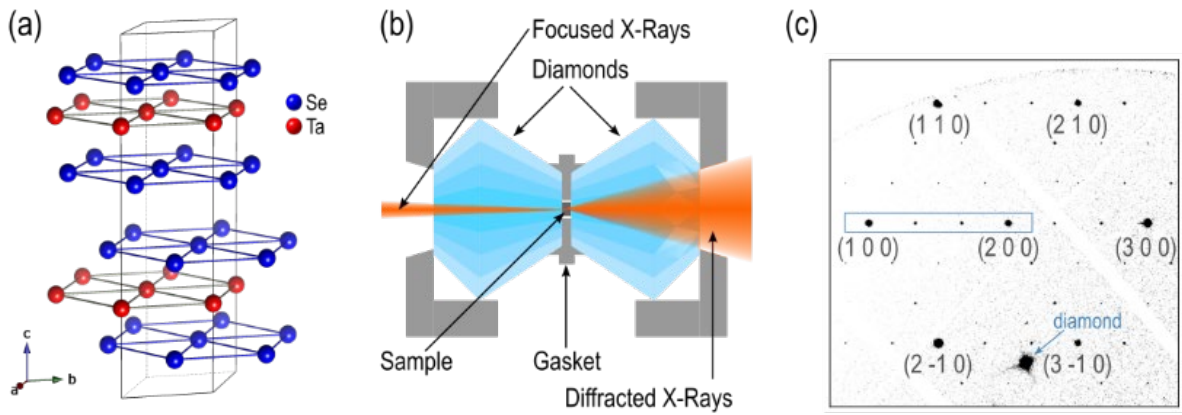

**FIG. 2**. (a) Hexagonal high-temperature structure of $2H$-$TaSe_2$ ($P6_3/mmc$, #194). Lines indicate the unit cell ($a = b = 3.425$ Å, $c = 12.57$ Å). (b) Schematics of a diamond anvil cell and scattering geometry used for the HP-LT XRD experiments. (c) Observed x-ray diffraction patterns for a $2H$-$TaSe_2$ single crystal at T = 40 K and p = 0.3 GPa shown for a ($HK0$) plane. The blue rectangle highlights the section for which results of a more detailed analysis are shown in Figure 4. The blue arrow indicates spurious Bragg scattering from the diamond single crystals.

|  | hexagonal | hexagonal | hexagonal$_{CDW}$ | hexagonal | hexagonal |
|---|---|---|---|---|---|
| Pressure (GPa) | 0.3 | 0.3 | 0.3 | 15.3 | 15.5 |
| Temperature (K) | 150 | 40 | 40 | 40 | 80 |
| Space Group | $P6_3/mmc$ | | | | |
| $a$ (Å) | 3.42621(9) | 3.42441(9) | 10.27599(14) | 3.2961(2) | 3.29460(15) |
| $b$ (Å) | 3.42621(9) | 3.42441(9) | 10.27599(14) | 3.2961(2) | 3.29460(15) |
| $c$ (Å) | 12.579(7) | 12.574(7) | 12.569(3) | 11.383(14) | 11.362(9) |
| $V^3$(Å$^3$) | 127.88(7) | 127.70(7) | 1149.4(3) | 107.10(14) | 106.81(9) |
| $\alpha$ (°) | 90 | | | | |
| $\beta$ (°) | 90 | | | | |
| $\gamma$ (°) | 120 | | | | |
| $R_{int}$ (%) | 1.9 | 3.1 | 2.3 | 2.4 | 5.2 |
| $R_1/wR_2$ (%) | 1.36/1.55 | 1.44/1.54 | 2.17/2.63 | 1.84/1.88 | 3.78/4.77 |

**TABLE 1**. Overview of the unit cell, space group, lattice parameters ($a$, $b$, and $c$), angles ($\alpha$, $\beta$, and $\gamma$), and $R$ factors for a representative set of pressure-temperature values. The thick rectangle denotes a refinement taking into account the CDW 3×3 supercell at low pressure. The other refinements were done disregarding the superlattice peaks. The detailed pressure dependence of the $z$ parameter (Se place) for the refinements of the high-temperature unit cell is displayed in Figure 4(d). All $x,y,z$ parameters (Se 4f Wyckoff position) for the supercell, crystallographic information files (CIF) for the refinements shown here and a more detailed table for all measured data sets can be found in the KITopen data repository[54].

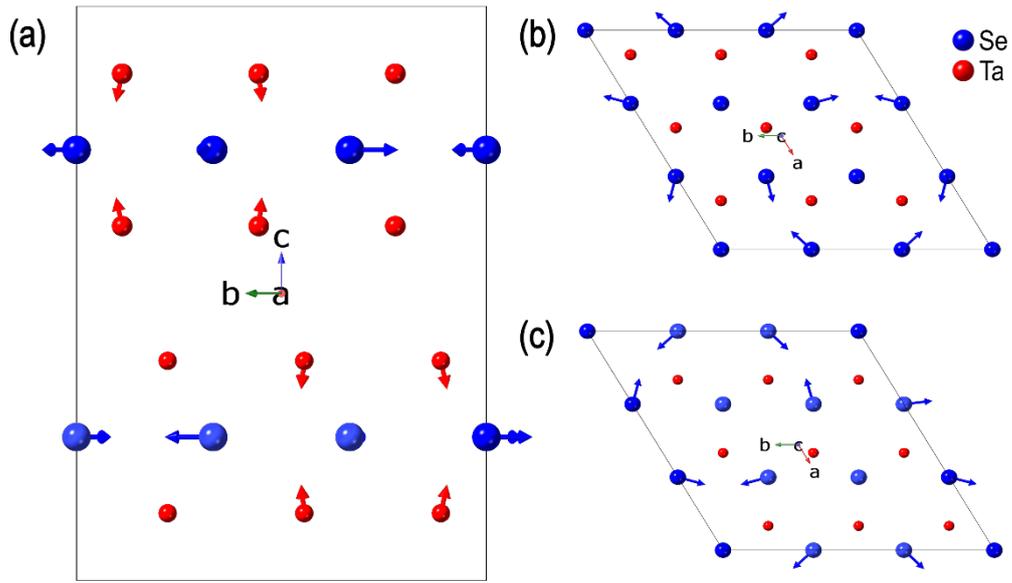

**FIG. 3**. Refined commensurate CDW structure at T = 40 K and p = 0.3 GPa. (a) Blue and red spheres indicate the positions of Ta and Se ions, respectively, in an undistorted 3×3×1 supercell of the high-temperature structure for a view along the *a* axis. The arrows indicate the displacement of each ion in the supercell as determined by our refinement. The Ta displacements were upscaled by a factor of 10 with regard to those of Se for visibility. (b) View along *c* axis for upper TaSe$_2$ sandwich layer. (c) View along *c* axis for lower TaSe$_2$ sandwich layer.

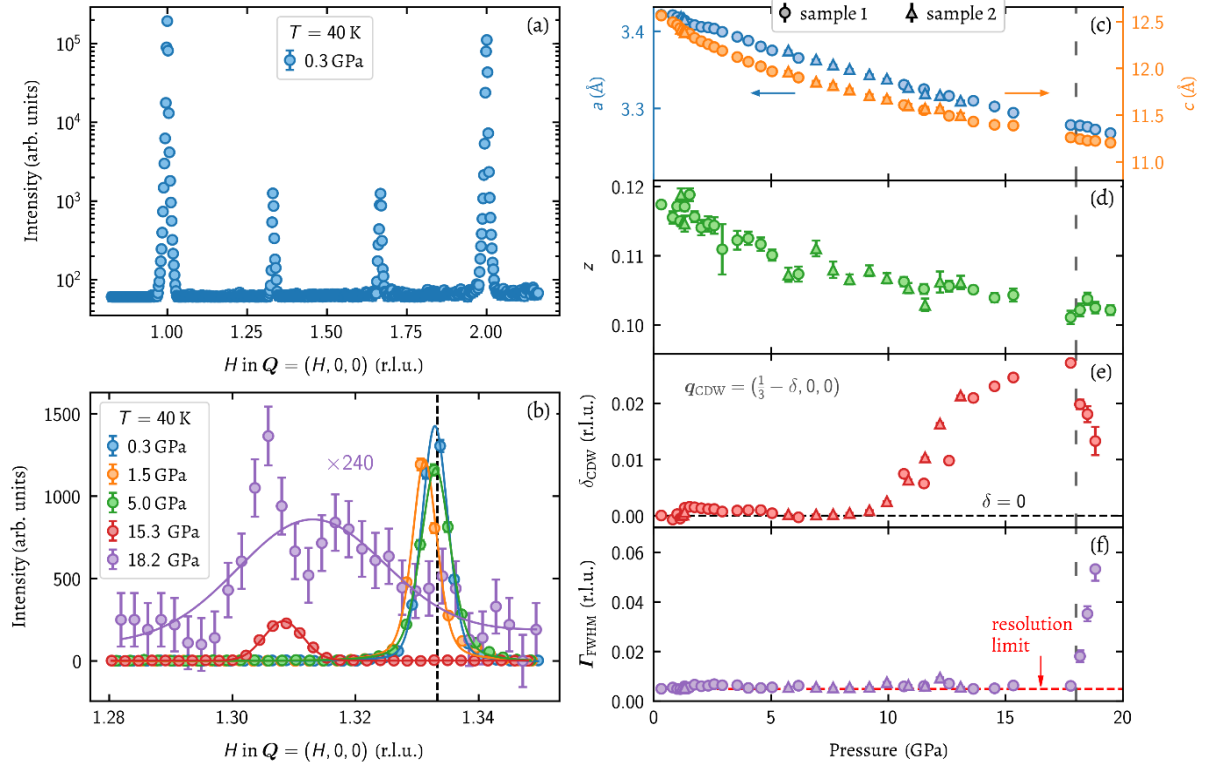

**FIG. 4**. Detailed pressure dependence at T = 40 K. (a) Line scan extracted for $Q = (1,0,0) - (2,0,0)$ [blue box in Fig. 2(b)]. Note that intensities are shown on a logarithmic scale. (b) Line scans through the CDW superlattice peak at $Q \approx (1.33,0,0)$ for T = 40 K and various pressures p = 0.3 GPa – 18.2 GPa. A constant background was subtracted. Data for 18.2 GPa was multiplied by a factor of 240 for visibility. The vertical dashed line denotes the position of the commensurate CDW, $q_{CCDW} = (1/3,0,0)$. Data at p = 18.2 GPa were taken with a higher flux than those at lower pressures (see text for details). (c) Lattice parameters (P6$_3$/mmc), (d) $z$ parameter (of Se atomic position in high-temperature unit cell), (e) incommensurability δ [defined by the superlattice peak position at $q_{CDW} = (1/3-\delta,0,0)$] and (f) linewidth of the superlattice peak $\Gamma_{FWHM}$ shown as a function of pressure for T = 40 K. The horizontal red dashed line in (f) denotes the measured widths of the nearby fundamental reflection, $Q^{Bragg} = (1,0,0)$ and, thus, represents the resolution limit. The vertical dashed line denotes the critical pressure at T = 40 K, $p_{c,T=40K} \approx 18$ GPa, defined by the increase of $\Gamma_{FWHM}$ to above the resolution limit.

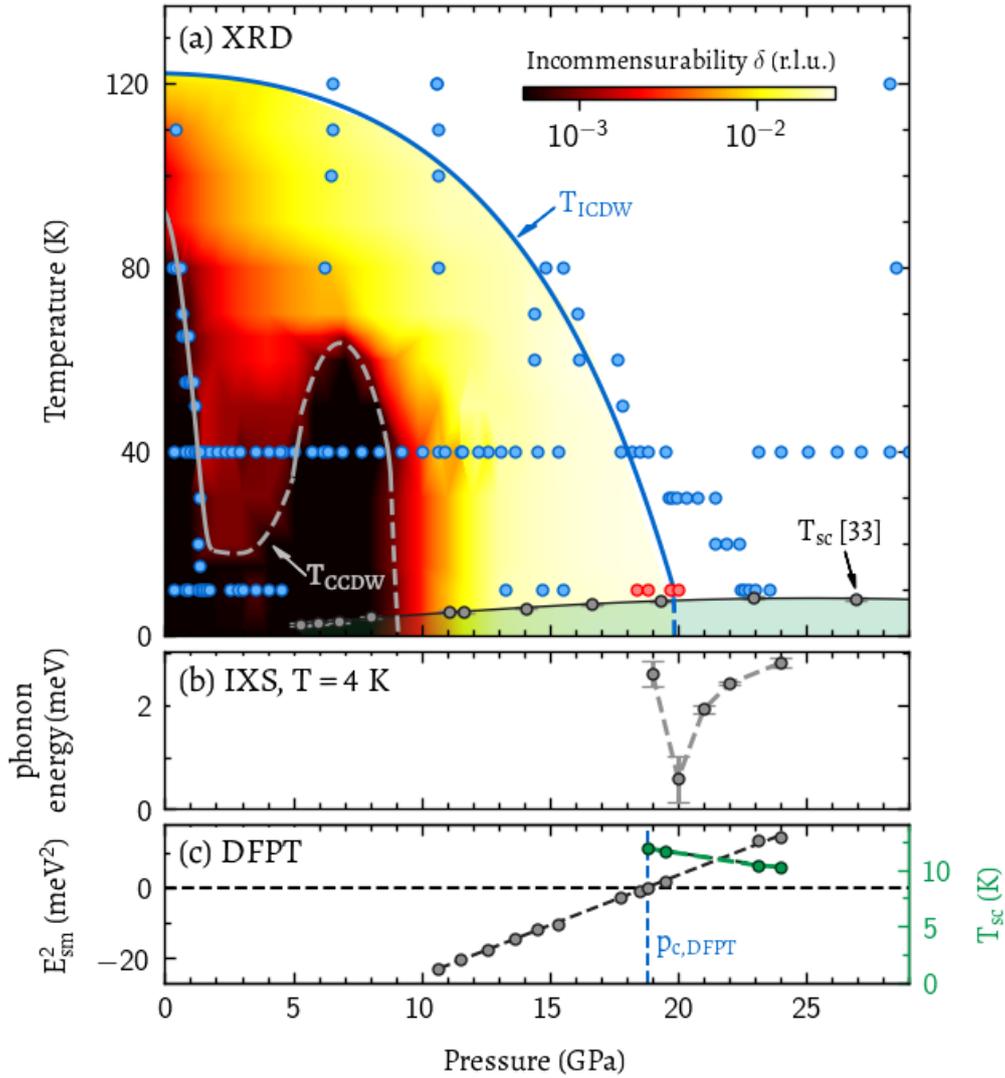

**FIG. 5**. (a) Temperature-pressure phase diagram of 2$H$-TaSe$_2$ derived from x-ray diffraction experiments. Blue (red) dots indicate the (T,p) points at which XRD (elastic IXS) measurements were performed. The blue/grey solid lines indicate guides to the eye for the pressure dependence of the $T_{ICDW}$/$T_{CCDW}$ crossover. The grey dashed line shows a possible evolution of $T_{CCDW}$ in regions of the phase diagram without measurements. The color-code reflects the deduced values of the incommensurability $\delta$ of the CDW ordering wave vector $\mathbf{q}_{CDW}$ = (⅓-$\delta$,0,0). Open circles represent the superconducting transition temperature $T_{sc}$ reported in Ref. [33]. (b) Energy of the CDW soft phonon mode deduced from our high-pressure IXS measurements at T = 4 K (see SI and Fig. S5 for details). (c) Calculated square phonon energy of the soft phonon mode in 2$H$-TaSe$_2$ based on DFPT (black symbols, left-hand scale). Pressure effects were included using the experimental lattice parameters observed in XRD at T = 40 K. The vertical dashed (blue) line indicates the pressure $p_{c,DFPT}$ = 18.8 GPa, for which a linear fit of the square phonon energy (dashed line) crosses zero. Calculations including electron-phonon coupling for p > $p_{c,DFPT}$ predict a superconducting $T_{sc,DFPT}$ = 11.8 K just above $p_{c,DFPT}$ which decreases for higher pressures (green symbols, right-hand scale).

# Supplemental Information to:

# Charge-density-wave quantum critical point under pressure in 2$H$-TaSe$_2$

Yuliia Tymoshenko,[1] Amir-Abbas Haghighirad,[1] Rolf Heid,[1] Tom Lacmann,[1] Alsu Ivashko,[1] Adrian Merritt,[2] Xingchen Shen,[1,3] Michael Merz,[1] Gaston Garbarino,[4] Luigi Paolosini,[4] Alexei Bosak,[4] Florian K. Diekmann,[5,6] Kai Rossnagel,[5,6,7] Stephan Rosenkranz,[8] Ayman H. Said,[9] and Frank Weber[1,†]

[1] Institute for Quantum Materials and Technologies, Karlsruhe Institute of Technology, 76131 Karlsruhe, Germany
[2] Physikalisches Institut, Karlsruhe Institute of Technology, 76131 Karlsruhe, Germany
[3] Laboratoire de Cristallographie et Sciences des Matériaux (CRISMAT), CNRS, ENSICAEN, Caen 14000, France
[4] ESRF, The European Synchrotron, 71 Avenue des Martyrs, CS40220, 38043 Grenoble Cedex 9, France
[5] Institute of Experimental and Applied Physics and KiNSIS, Kiel University, 24098 Kiel, Germany
[6] Ruprecht Haensel Laboratory, Kiel University, 24098 Kiel, Germany
[7] Ruprecht Haensel Laboratory, Deutsches Elektronen-Synchrotron DESY, 22607 Hamburg, Germany
[8] Material Science Division, Argonne National Laboratory, Lemont (IL), USA
[9] Advanced Photon Source, Argonne National Laboratory, Lemont (IL), USA

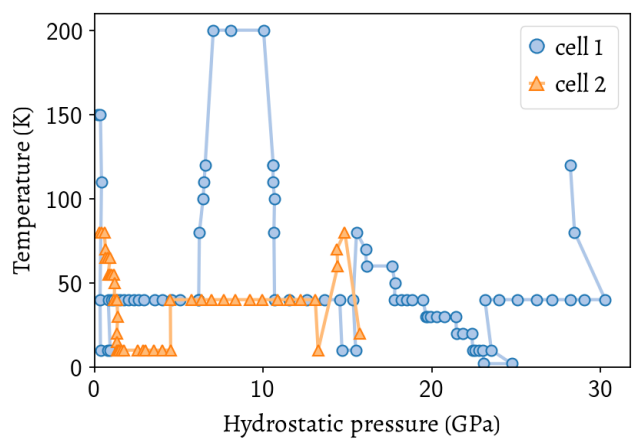

**Fig. S1.** Sequence of temperature-pressure points measured with XRD at beamline ID15b for DAC 1 (circles) and DAC 2 (triangles).

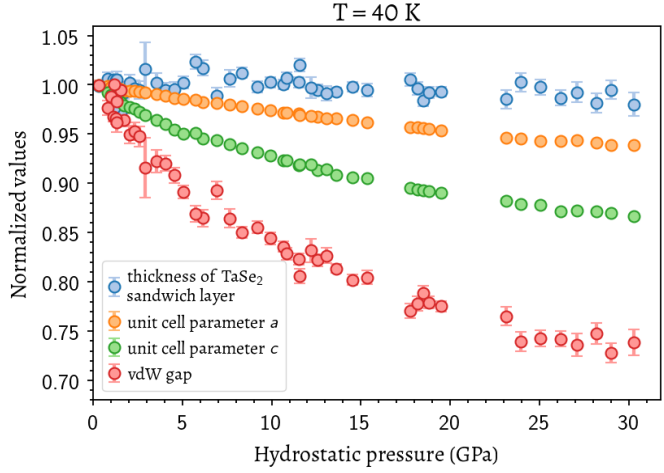

**Fig. S2.** Pressure dependence of various structural parameters as deduced in our structural refinements for the high-temperature unit cell at T = 40 K. All parameters are normalized to their respective value at the lowest pressure investigated, $p_{min}$ = 0.3 GPa. The size of the van-der-Waals (vdW) gap and the thickness $d$ of a single TaSe$_2$ sandwich layer are derived from the $z$ parameter of the Se atomic site [see Fig. 4(d)] and the c axis lattice parameter [see Fig. 4(c)] according to:

$$vdW = 2 * c * z$$

$$d = 0.5 * c * (1 - 4 * z)$$

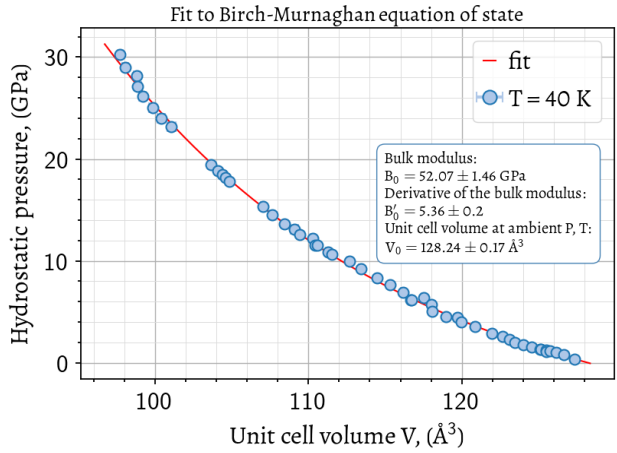

**Fig. S3.** Birch-Murnaghan fit of the pressure dependent unit cell volume of 2H-TaSe$_2$ (see equation below). The resulting parameters are given in the legend. The fit was performed using the third-order Birch-Murnaghan equation of state (see equation 1; providing insights into the material's compressibility and structural stability), where $p$ = pressure, $V$ = volume at a given pressure, $V_0$ = volume at zero pressure, $B_0$ = Bulk modulus at zero pressure and $B_0'$ = Derivative of the bulk modulus with respect to pressure.

$$p(V) = \frac{3B_0}{2}\left[\left(\frac{V_0}{V}\right)^{7/3} - \left(\frac{V_0}{V}\right)^{5/3}\right]\left\{1 + \frac{3}{4}(B_0' - 4)\left[\left(\frac{V_0}{V}\right)^{2/3} - 1\right]\right\} \qquad (1)$$

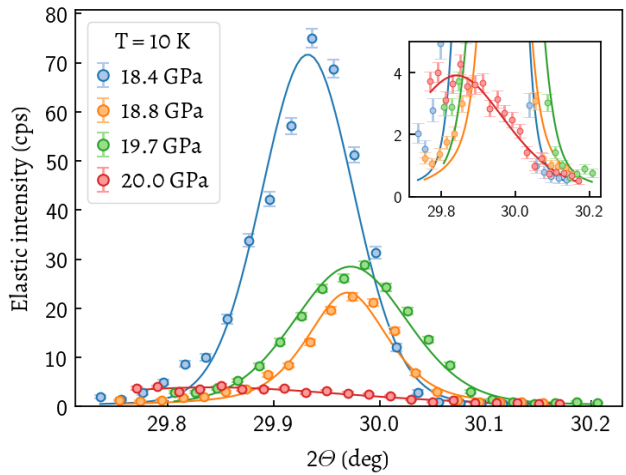

**Fig. S4.** Scattering intensity of elastic scans (zero energy transfer) through the CDW superlattice peak at $Q = (2.69, 0, 1)$ obtained on the ID28 HERIX spectrometer at ESRF, France. The inset focuses on the low-intensity observed at p = 20 GPa.

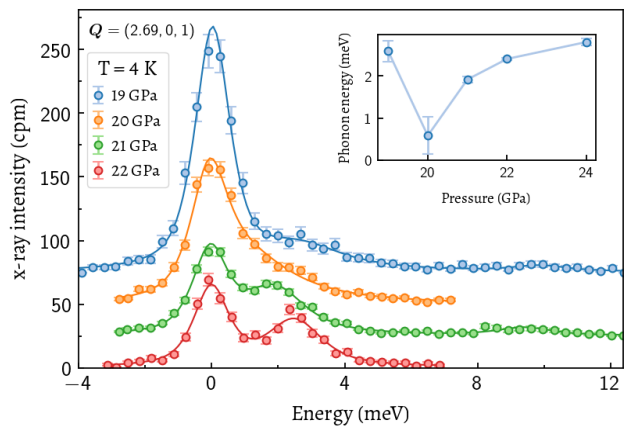

**Fig. S5.** Energy scans at $Q = (2.69, 0, 1)$ obtained on the ID28 HERIX spectrometer at ESRF, France, for various pressures, p = 19 GPa – 22 GPa. Lines represent fits to the data consisting of a resolution-limited pseudo-Voigt function fixed to zero energy transfer and DHO functions, convoluted with the experimental resolution, for the phonons. The inset shows the pressure dependence of the energy of the CDW soft phonon mode for p = 19 GPa – 24 GPa.